\documentclass[12pt,twoside,letterpaper]{article}

\usepackage{graphicx}
\usepackage{multirow}
\usepackage{xspace}
\usepackage{boxedminipage}
\usepackage{epstopdf}

\usepackage{color}
\usepackage{eso-pic} 
\usepackage{ifpdf}
\ifpdf
 \usepackage{url,boxedminipage}
 \usepackage{graphicx,thumbpdf}
 \definecolor{rltred}{rgb}{0.75,0,0}
 \definecolor{rltgreen}{rgb}{0,0.5,0}
 \definecolor{rltblue}{rgb}{0,0,0.75}
 \DeclareGraphicsExtensions{.pdf,.jpg,.jpeg}            
\else
 \usepackage{graphicx}                                  
 \DeclareGraphicsExtensions{.eps,.ps,.eps.gz,.ps.gz}    
 \newcommand{\href}[2]{#2}

\fi 

\usepackage{multirow}

\def\hbb{$H \to b\bar{b}$\xspace}
\def\hgg{$H \to \gamma\gamma$\xspace}
\def\hzz{$H \to Z Z$\xspace}
\def\hww{$H \to W^+W^-$\xspace}

\def\met{$\not\!\!E_T$\xspace}
\def\mjj{$m_{jj}$\xspace}

\begin{document}
\thispagestyle{empty}
\vspace*{-3.5cm}
\begin{flushright}
CDF/PUB/ELECTROWEAK/PUBLIC/10860\\
Version 2.0\\
\today
\end{flushright}

\vspace{0.5in}

\begin{center}
  \begin{large}
  {\bf Standard Model Higgs Searches at the Tevatron}
  \end{large}
\end{center}


\begin{center}
\href{mailto:knoepfel@fnal.gov}{Kyle.~J.~Knoepfel}\footnote{E-mail: {\tt knoepfel@fnal.gov}}\textsuperscript{,}~
\footnote{Contributed to the Proceedings of the XX International Workshop on Deep-Inelastic Scattering and Related Subjects, March 2012, Bonn, Germany}\\ 
(\emph{On behalf of the CDF and D0 collaborations})\\ \vspace{1.0ex} 
\emph{Fermi National Accelerator Laboratory}
\end{center}

\begin{abstract}
	We present results from the search for a standard model Higgs
	boson using data corresponding up to 10 fb${}^{-1}$ of
	proton-antiproton collision data produced by the Fermilab
	Tevatron at a center-of-mass energy of 1.96 TeV. The data were
	recorded by the CDF and D0 Detectors between March 2001 and
	September of 2011.  A broad excess is observed between $105 <
	m_H < 145 \ \mathrm{GeV}/c^2$ with a global significance of
	2.2 standard deviations relative to the background-only
	hypothesis.
\end{abstract}

\section{Introduction}

The Higgs boson is the only standard model (SM) particle yet to be
found.  Within the SM, it is the particle responsible for giving the
$W$ and $Z$ bosons and fermions their masses~\cite{higgstheory}.
Although the value of the Higgs mass is a parameter, and is therefore
formally unconstrained, measurements of electroweak observables such
as $m_W$ and $m_t$, indirectly constrain it to be less than roughly
150 GeV$/c^2$ at 95\% confidence level (C.L.).

In December of 2011, the CMS and ATLAS collaborations released results
indicating excesses in the \hgg channel relative to the
background-only hypothesis.  Along with the \hgg channel, the \hww and
\hzz diboson channels have the most sensitivity to Higgs
production~\cite{cmshiggs,atlashiggs}.  On the other hand, the
sensitivity to the Higgs through the \hbb channel is a bit less --- as
of the Moriond 2012 conference, the expected sensitivities to the \hbb
final states are roughly $4.3\times \sigma_\mathrm{SM}$~\cite{cmshbb}
and $3.5\times \sigma_\mathrm{SM}$~\cite{atlashbb}, respectively, for
the CMS and ATLAS Higgs searches. In contrast, the expected
sensitivities at the Tevatron at low-mass are predominantly through
the \hbb associated production modes, and on the order of
$1.7\times\sigma_\mathrm{SM}$~\cite{cdfhiggs} and
$2.2\times\sigma_\mathrm{SM}$~\cite{d0higgs}, respectively, for the
CDF and D0 searches.  The low-mass Higgs searches at the Tevatron and
LHC are therefore complementary.  The \hbb searches at the Tevatron
are therefore the focus of this presentation.

\section{Low-Mass Higgs Searches at the Tevatron}

Low-Mass Higgs searches use a typical analysis selection, summarized
in Table~\ref{table:sel}.  Requiring the two leading jets in the event
to be $b$-tagged substantially reduces the background relative to the
Higgs signal.  Additional sensitivity can be gained by also
considering events where one but not both leading jets are tagged.
Note that analyses that have zero and one leptons in the final state
are also sensitive to $\ell\nu b\bar{b}$ and $\ell^+\ell^-b\bar{b}$
final states, respectively, due to losing a lepton from detector
inefficiencies.  Multivariate techniques are implemented to separate
signal from QCD and electroweak backgrounds.

\begin{table}
  \centering
  \begin{tabular}{lccc} \hline \hline
    \multirow{2}{*}{Analysis Channel}& No. of & \multirow{2}{*}{\met} & No. of \\
    & Leptons & & $b$-jets	\\ \hline
    $ZH \to \nu\bar{\nu}b\bar{b}$ & 0 & Yes & 2 \\
    $WH \to \ell\nu b\bar{b}$ & 1 & Yes & 2 \\
    $ZH \to \ell^+\ell^-b\bar{b}$ & 2 & No & 2 \\ 
    \hline \hline
  \end{tabular}	
  \caption{Basic selection for \hbb final states  produced in association with a $W$ or $Z$ boson.}
  \label{table:sel}
\end{table}

To obtain the best expected sensitivities to SM Higgs production, the
$b$-tagging and lepton identification algorithms must be optimized.
Also, variables that discriminate between signal and background
processes can be optimized or improved, such as the dijet invariant
mass, which improves the ability of multivariate discriminants to
separate the Higgs signal from background processes.

\subsection{Analysis Improvements}

The results presented here are given in Ref.~\cite{higgscombo} and the
references therein.  Many Higgs analyses at CDF and D0 have
implemented several improvements, consisting of increased luminosity
(roughly 10\% gain in sensitivity), improved $b$-tagging, improved
\mjj resolution, and various improvements in analysis methods.  At
CDF, an improved $b$-tagging algorithm called HOBIT~\cite{hobit} was
used in most of the mainstream \hbb search channels.  This
multivariate algorithm, trained on \hbb events for a Higgs mass of
$m_H = 120 \ \mathrm{GeV}/c^2$, increased Higgs sensitivity by roughly
10\% for a given search channel.  Various analysis improvements were
made by the D0 searches, including increasing signal acceptance by
relaxing some of the variable definitions.

\section{Tevatron Combinations}

To determine upper limits on SM Higgs production at 95\% C.L., two
approaches are taken: a frequentist profile likelihood approach where
the maximum of the likelihood is used to determine the nuisance
parameters; a modified frequentist (or Bayesian) approach where the
nuisance parameters are integrated out to determine posterior
probabilities.  Both approaches yield agreement better than 10\% for
all Higgs mass hypotheses in the range $100 < m_H < 200\
\mathrm{GeV}/c^2$, and better than 1\% on average.  The Bayesian
limits are adopted as the Tevatron results.

Systematic uncertainties such as the jet-energy scale, luminosity, and
pdf's are incorporated as nuisance parameters.  Care is taken to
ensure that all appropriate correlations are taken into account
between the CDF and D0 experiments.

\subsection{All Channels}

\begin{figure}[htb]
\includegraphics[width=0.48\textwidth]{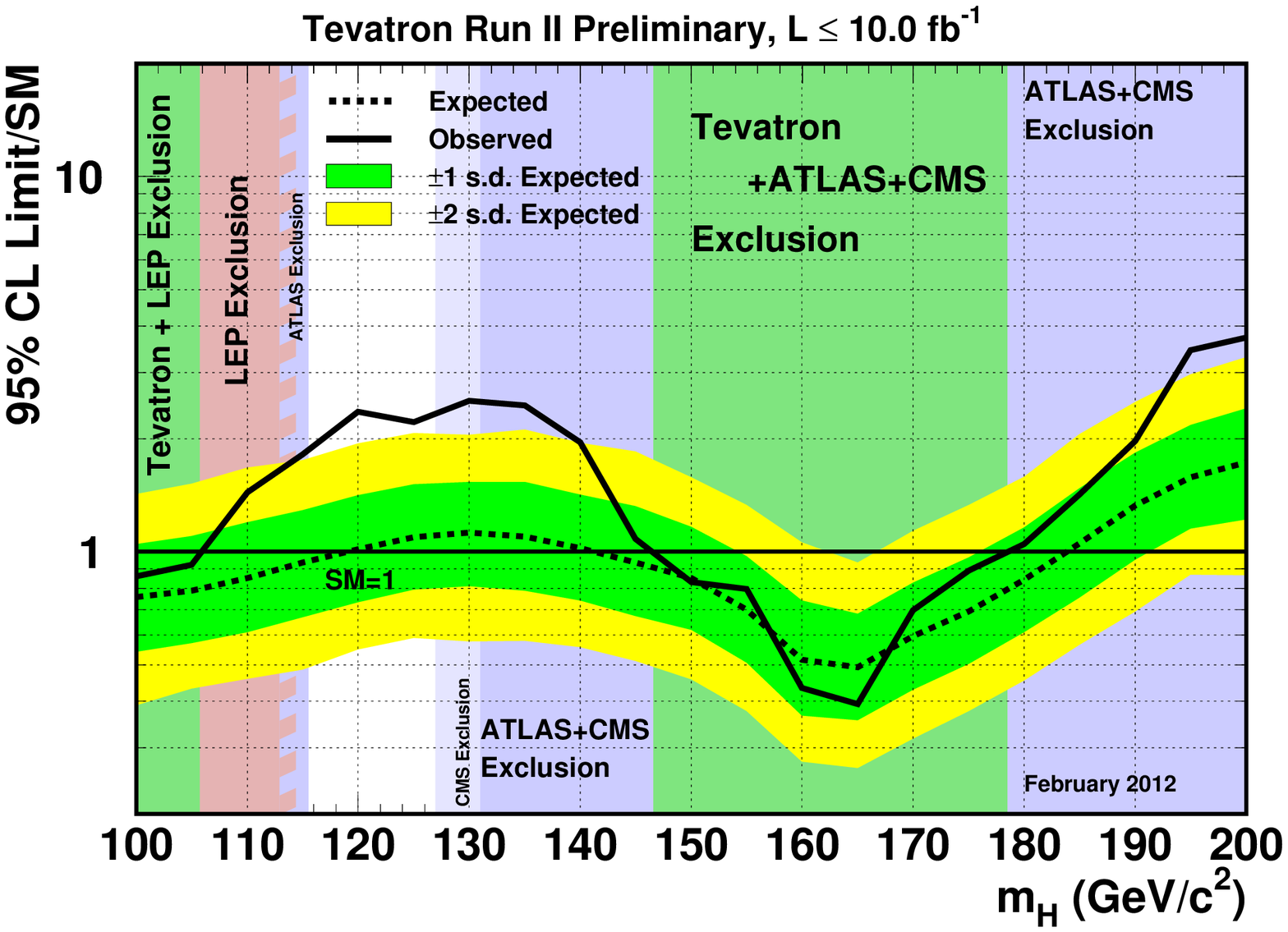} \hfill
\includegraphics[width=0.48\textwidth]{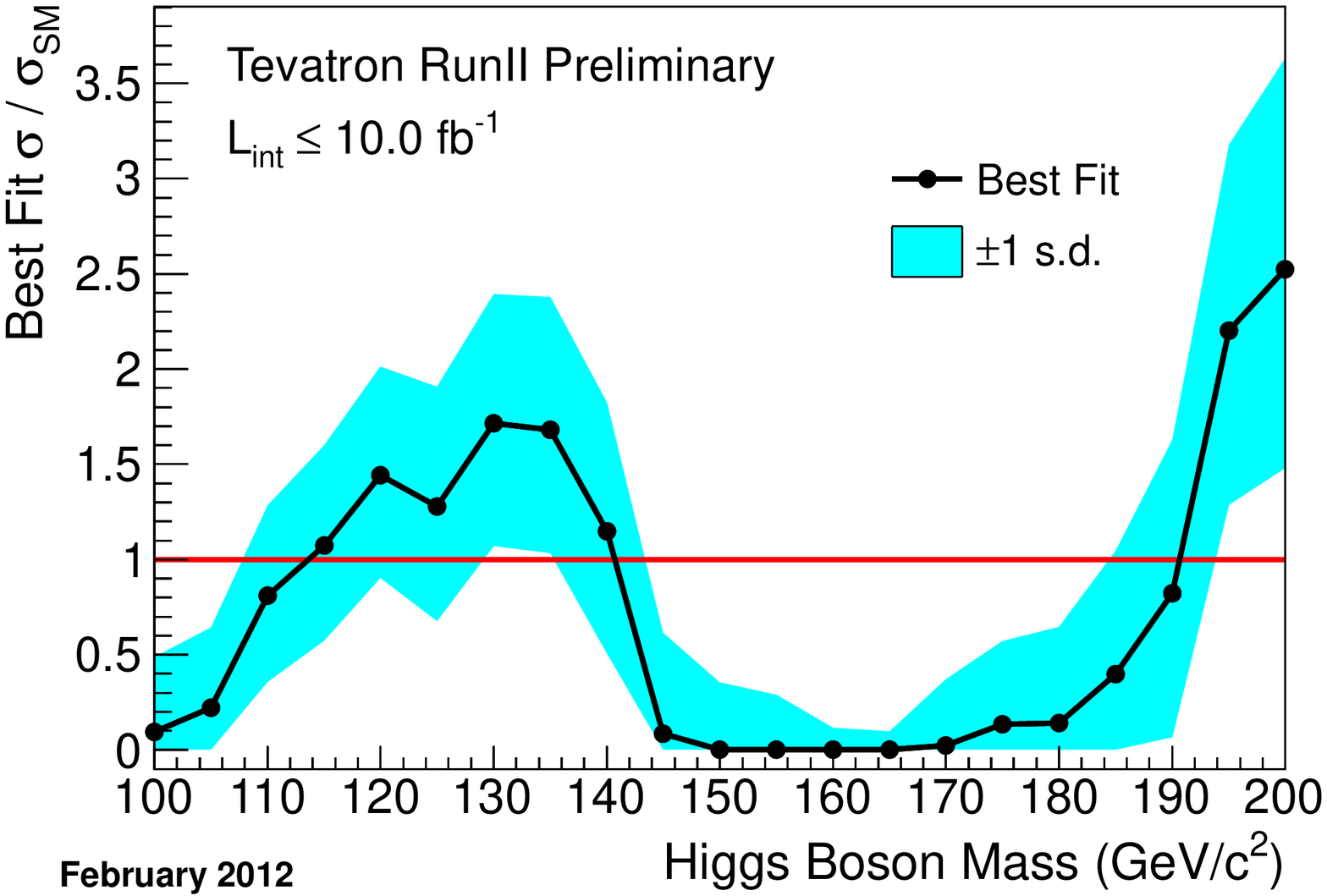}
\caption{(Left) Upper limits on SM Higgs production at 95\% C.L.
assuming the background-only hypothesis.  (Right) Best fit value for
SM Higgs production, assuming the signal plus background hypothesis.}
\label{fig:smhiggslimits}
\end{figure}

The complete list of channels that goes into the Higgs Tevatron
combination is given in Ref.~\cite{higgscombo}, along with a complete
description of the limit-setting procedure and handling of systematic
uncertainties.  The 95\% C.L. limits on Higgs production are shown in
Fig.~\ref{fig:smhiggslimits}, along with the best-fit value assuming
the existence of signal in the data.  The expected upper limit is
better than $1.15\times\sigma_\mathrm{SM}$ across the entire Higgs
mass range.  The expected exclusion regions are $100 < m_H < 119\
\mathrm{GeV}/c^2$ and $141 < m_H < 184\ \mathrm{GeV}/c^2$, whereas the
observed exclusion regions are $100 < m_H < 106\ \mathrm{GeV}/c^2$ and
$147 < m_H < 179\ \mathrm{GeV}/c^2$.  A broad excess is observed in
the mass range $105 < m_H < 145\ \mathrm{GeV}/c^2$, where the minimum
local $p$-value ($p_0$) at $m_H = 120\ \mathrm{GeV}/c^2$ corresponds
to a 2.7-standard-deviation departure from the background-only
hypothesis.  Assuming a jet-energy resolution of about 10\%, the
estimated correction from the look-elsewhere effect gives
$p_\mathrm{global} \approx 4\times p_0$, which corresponds to a $2.2$
standard-deviation effect relative to the background-only hypothesis.
The best-fit value of the Higgs production cross-section is consistent
with the SM prediction in the mass region $110 < m_H < 140\
\mathrm{GeV}/c^2$.

\subsection{\hbb Channels}

\begin{figure}[htb]
\includegraphics[width=0.48\textwidth]{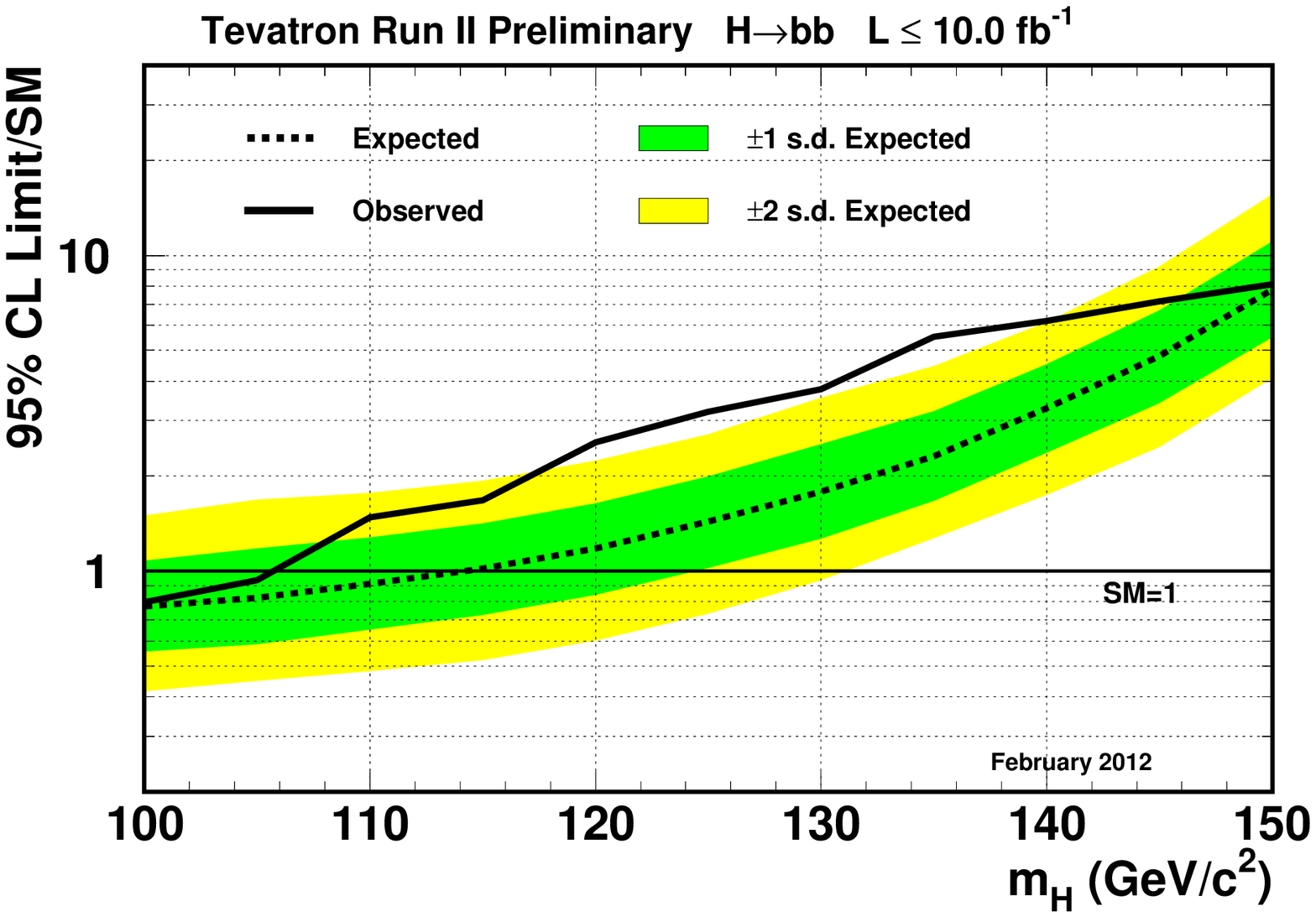} \hfill
\includegraphics[width=0.48\textwidth]{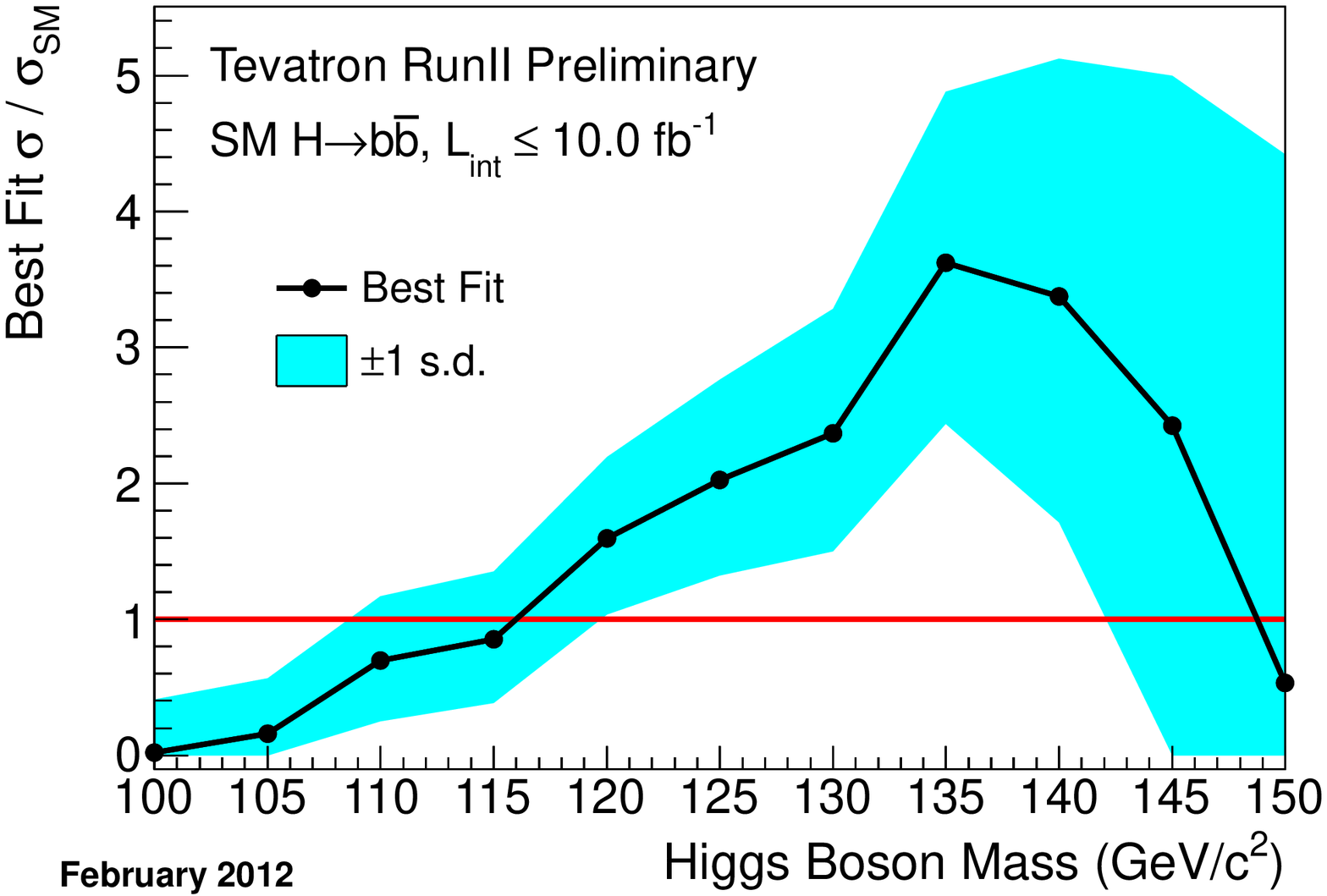}
\caption{(Left) Upper limits on \hbb production at 95\% C.L.  assuming the background-only hypothesis.  (Right) Best fit value
for \hbb production, assuming the signal plus background hypothesis.}
\label{fig:hbblimits}
\end{figure}

Upper limits on SM Higgs production have also been obtained by looking
at only \hbb final states.  The 95\% C.L. limits, and associated
best-fit value for Higgs production are shown in
Fig.~\ref{fig:hbblimits}.  The broad excess in the low-mass range
results in a minimum $p$-value of 2.8 standard deviations away from
the background-only hypothesis at a Higgs mass of $m_H = 135\
\mathrm{GeV}/c^2$.  Using a look-elsewhere effect correction of 2
(based on the assumed jet-energy resolution), the global $p$-value is
diluted to a 2.6 standard-deviation effect.  The best-fit value is
consistent with the SM prediction in the region $110 < m_H < 120\
\mathrm{GeV}/c^2$.  The significant departure from the SM prediction
in the region $120 < m_H < 145\ \mathrm{GeV}/c^2$ results from the
signal component of the fit needing to compensate for the excess of
data events relative not only to the background-only prediction, but
also to the signal-plus-background prediction.

\section{Conclusions}

The CDF and D0 Collaborations have combined their results to give a
Tevatron-wide combination of the upper limits of the SM Higgs
production at 95\% C.L.  After combining all channels across the range
$100 < m_H < 200\ \mathrm{GeV}/c^2$, a broad excess is observed in
data relative to the background-only hypothesis, corresponding to a
2.2 standard-deviation effect.  Considering only the \hbb search
channels yields an excess in data, corresponding to a 2.6
standard-deviation departure from the background-only prediction.  An
update of the Tevatron combination is expected in the summer of 2012.

\section{Acknowledgments}

The author wishes to thank the DIS workshop organizers and conveners
for an enjoyable and rewarding experience.  He acknowledges support
from Fermi National Accelerator Laboratory and the CDF experiment.

{\raggedright
\begin{footnotesize}

\end{footnotesize}
}


\end{document}